\documentclass[twocolumn, amsmath, amssymb]{revtex4-1}

\usepackage[utf8x]{inputenc}
\usepackage[T1]{fontenc}
\usepackage{bm}
\usepackage{graphicx}
\graphicspath{{./figures_sources/}}
\usepackage{tikz}
\usepackage{lipsum}
\usetikzlibrary{arrows, snakes, backgrounds, plotmarks, patterns, external}
\usepackage{pgfplots}

\newlength
\figureheight
\newlength
\figurewidth 

\definecolor{clr1}{RGB}{40,56,120}		
\definecolor{clr2}{RGB}{76,166,107}		
\definecolor{clr3}{RGB}{228,66,45}		
\definecolor{slabback}{RGB}{180,180,180}
\definecolor{clr4}{rgb}{0,0.4470,0.7410}
\definecolor{clr5}{rgb}{0.8500,0.3250,0.0980}
\definecolor{clr6}{rgb}{0.9290,0.6940,0.1250}
\definecolor{clr7}{rgb}{0.4940,0.1840,0.5560}
\definecolor{clr8}{rgb}{0.4660,0.6740,0.1880}
\definecolor{clr9}{rgb}{0.3010,0.7450,0.9330}
\definecolor{clr10}{rgb}{0.6350,0.0780,0.1840}
\definecolor{clr11}{rgb}{0.102,0,0.51}
\definecolor{clr12}{rgb}{0.12156862745098,0.466666666666667,0.705882352941177}

\newcommand{\avrg}[1]{\langle#1\rangle}

\renewcommand{\vec}[1]{\mathbf{#1}}

\def\tmfp{\ell}				
\def\rL{s}					
\def\g{g}		
\def\gbrouwer{g_\textup{B}}	
\def\s_factor{\chi}					
\def\Tg{T}		
\def\Lslab{L_s}		
\def\ks{q}		
\def\k{k}		

\def\tmfp{\ell}				
\def\rL{s}					
\def\bal{\ell_\textup{a}}		
\def\rdal{s_\textup{a}}		

\begin{document}


\title{
Experimental characterization of rigid scatterer hyperuniform distributions for audible acoustics
}

\author{Elie Chéron}
\email{elie.cheron@univ-lemans.fr}
\author{Jean-Philippe Groby}
\author{Vincent Pagneux}
\author{Simon Félix}
\author{Vicent Romero-García}
\affiliation{Laboratoire d’Acoustique de l’Université du Mans (LAUM), UMR 6613, Institut d'Acoustique - Graduate School (IA-GS), CNRS, Le Mans Université, France}

\date{\today}

\begin{abstract}
Two-dimensional stealthy hyperuniform distributions of rigid scatterers embedded in a waveguide are experimentally characterized the wave transport properties for scalar waves in airborne audible acoustics. The non resonant nature of the scatterers allows us to directly links the these properties to the geometric distribution of points through the structure factor. The transport properties are analyzed as a function of the stealthiness $\chi$ of their hyperuniform point pattern and compared to those of a disordered material in the diffusive regime, which are characterized by the Ohm’s law through the mean free path. Different scattering regimes are theoretically and numerically identified showing transparent regions, isotropic band gaps, and anisotropic scattering depending on $\chi$. The robustness of these scattering regimes to losses which are unavoidable in audible acoustics is experimentally unvealed. 
\end{abstract}

\maketitle

\section{Introduction}
Wave transport properties in complex systems are one of the most studied topics in wave physics. Waves traveling in these complex systems often undergo multiple scattering. Depending on both the distribution of scatterers and the physical properties of each scatterer, several phenomena can appear in different ranges of frequencies as for example Anderson localization \cite{Schwartz07,Wiersma} in disordered systems or wave collimation \cite{Lu06} and focusing \cite{Luo02} in periodic systems. The opening of band gaps is probably the most celebrated phenomenon and has given rise to a plethora of studies in electromagnetic \cite{joannopoulos08} or elastic \cite{Deymier13} wave transport. In photonic \cite{Yablonovitch87, John87} or phononic \cite{Sigalas92, Kushwaha93} crystals, two main phenomena contribute to the generation of the band gaps \cite{Lidorikis98, Lidorikis00}: the Bragg scattering (based on geometrical arguments) and the excitation of the single scatterer Mie resonances (based on intrinsic local properties of the scatterer) \cite{Mie08, Bohren83}. While Bragg scattering establishes the necessary condition for the opening of band gaps, Mie resonances are helpful for the opening of full band gaps in two or three dimensional systems \cite{Lidorikis00, Rockstuhl06, Amoah16}. 

Recently, stealthy hyperuniform materials have emerged as amorphous systems presenting unique wave transport properties due to the correlated disorder and therefore opening new venues for controlling waves \cite{Man13a, Man13b, Leseur16, Gkantzounis17, Aubry20, Rohfritsch20, Romero21}. Hyperuniform materials are 
based on the concept of hyperuniformity which enforces 
the suppression of the long-range density fluctuations of the point pattern \cite{Torquato03, Uche04, Batten08, Torquato15, Torquato16,Froufe16}. Stealthy hyperuniform point patterns are characterized by the stealthiness, $\chi$, which imposes constrains on the structure factor in the reciprocal space. Several works have shown the evolution of the point distribution with $\chi$ in 1D \cite{Fan91} and 2D \cite{Uche04} systems. Three classes of point patterns can be distinguished in function of $\chi$: disordered, wavy-crystalline, and crystalline. While the $\chi$ boundary values for each of these classes are clearly established in 1D systems \cite{Fan91}, they depend on the number of particles in 2D systems \cite{Uche04}. While an anisotropic structure factor in the reciprocal space is obtained for $\chi\gtrsim0.55$ \cite{Froufe16}. 
and isotropic structure factor in the reciprocal space is obtained for $\chi\lesssim0.45$ \cite{Froufe16}, i.e., with low angular fluctuations.

Wave transport properties in materials made of hyperuniform distributions of scatterers, i.e., hyperuniform materials, have recently been analyzed for electromagnetic \cite{Man13a, Man13b} and elastic \cite{Gkantzounis17} waves by combining the geometric properties of stealthy hyperuniform point patters with the Mie resonances of the scatterers. So designed hyperuniform materials are transparent to incident long-wavelength excitation and possessed isotropic band gap at shorter wavelengths with $\chi\lesssim0.45$. This last features is in opposition to the quasi-periodic systems in which anisotropic band gaps are created with $\chi\gtrsim0.5$ \cite{Froufe16, Froufe17}. Waveguides with arbitrary paths have been designed by exploiting these isotropic full band gaps, \cite{Man13a, Man13b}. Nevertheless, Mie resonances are always used in all these examples, although hyperuniform materials should be uniquely charactized by the spatial Fourier transform of the point pattern. When disorder is introduced into periodic systems Bragg resonances and Mie resonance together have been shown to play an important role in the wave transport properties \cite{Lidorikis98, Rockstuhl06}. The band gap closes rapidly, when it is due to Bragg scattering, while is more robust, when it is due to Mie resonances with increasing disorder \cite{Lidorikis00, Amoah16}. It is therefore natural to ask what is the role of the Mie resonances with respect to the hyperuniform arrangement.

\begin{figure}
	\centering
\includegraphics[width=8cm]{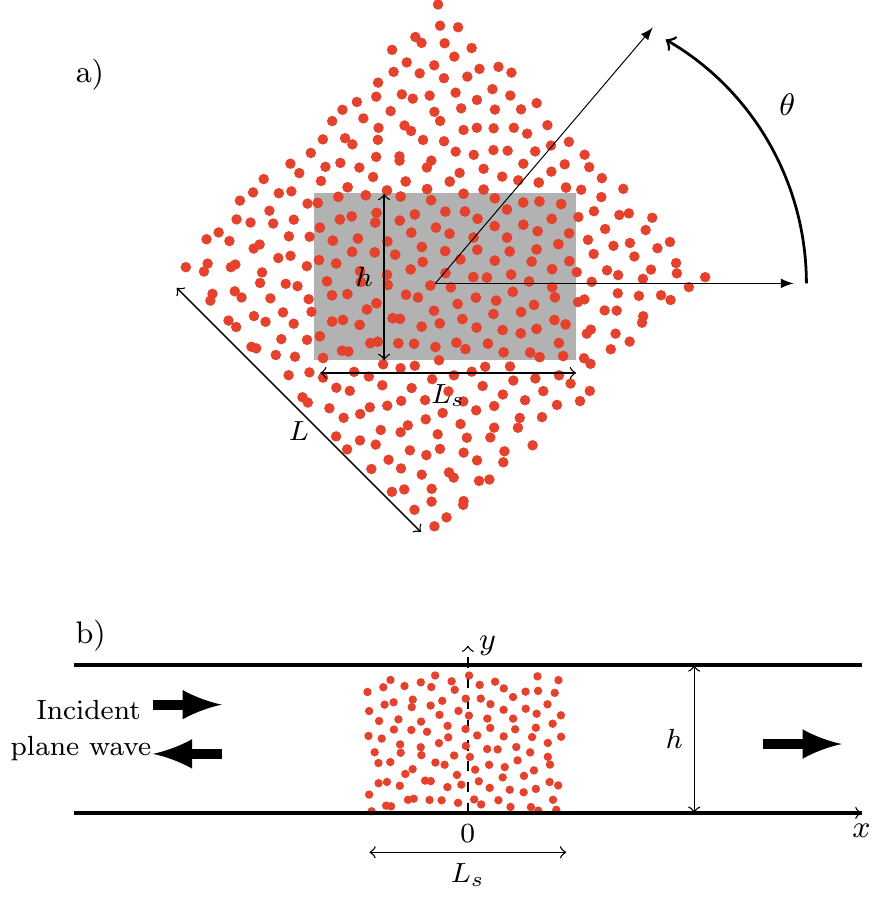}
	\caption{ (a) hyperuniform point pattern of $N=600$ points in a square region of side $L$. The point pattern can be rotated an angle $\theta$ with respect to its center. The point pattern has been generated by using the procedure described in the main text and in Ref.~\cite{Froufe16}. The point pattern shown in the area $L_s\times h$ delimited by the dashed lines is used to extract the point pattern that will be used to made a hyperuniform material by placing rigid scatterers at those positions. Inside this area the number of points will be $N_s\simeq 200$. We characterize the acoustic wave transport properties of such material by measuring the scattering coefficients of the sample when embedded in a rectangular acoustic waveguide. (b) Shows a picture of the acoustic waveguide used in the experimental set-up displaying how the material is placed inside. The material is excited by an incident plane wave. For each angle $\theta$, a different material can be created and then analyzed though the transmission coefficient as a function of the angle $\theta$.
	}
	\label{fig:fig1}
\end{figure}

In this work, two-dimensional (2D) hyperuniform distributions of rigid scatterers embedded in a rectangular cross-sectional waveguide are numerically and experimentally characterized for scalar waves in airborne audible acoustics. In this regime, solid scatterers are usually non penetrable (impervious) and present a Neumann boundary condition due to the huge impedance mismatch between their properties and those of the air medium \cite{bruneau_fundamentals_2006}. These scatterers are thus considered acoustically rigid and do not resonate. Therefore, airborne audible acoustics seems to be a good candidate to investigate connection between the wave transport properties of a given distribution of rigid scatterers and its corresponding structure factor. We perform a full wave solution of the Helmholtz equation to obtain the transport properties of the system made of a discrete distribution of rigid scatterers that can be compared with the structure factor result, and with experiments. Moreover, viscothermal losses are not avoidable in acoustics. As a consequence, we have solved the problem considering an absorption length in order to account for the viscothermal losses. The good agreement between experiments, simulations, and theory shows that the transport properties of the hyperuniform materials are robust to the presence of losses. We characterize the transition from a random to a periodic distribution by changing the stealthiness of the system, $\chi$. The transport properties for low values of $\chi$ (uncorrelated disorder) are well captured by the Ohm’s law using the theoretical mean free path of the media. As soon as the value of $\chi$ increases, the deviation of the transmission and the variance from the Ohm’s law at the Bragg frequency evidences the presence of the isotropic band gaps which becomes anisotropic for $\chi\geq0.5$. These results emphasize the relation of the transport properties in hyperuniform materials with the geometric distribution of points through the structure factor by eliminating the local resonance of the scatterers.

\section{Hyperuniform point patterns}
Let us consider a distribution of $N$ points located at positions $\bm{r_i}\;(i=1,...,N)$ inside a square domain of side $L$, as shown in Fig. \ref{fig:fig1}(a). The structure factor, $S(\vec{q})$, of this point pattern is defined as its spatial Fourier transform and reads as follows \cite{kittel04, Ashcroft}
\begin{equation}
S(\bm{\ks}) = \dfrac{1}{N}\sum\limits_{i=1}^{N}\sum\limits_{j=1}^{N} e^{-i\bm{\ks}.(\bm{r}_j-\bm{r}_i)},
\end{equation}
where $\bm{q}$ is a vector in the Fourier space. It is worth noting here that if the point pattern is periodic, the structure factor will present the characteristic Bragg peaks \cite{Froufe16}. For example, they are located at $\bm{q_B}=$~$(n2\pi\sqrt{N}/L, m2\pi\sqrt{N}/L)$, with $m, n\in\mathbb{Z}$ for a two-dimensional square array which periodicity is $a=L/\sqrt{N}$. For periodic structures, $\bm{q_B}$, represents the well-known vectors of the reciprocal lattice in the reciprocal space. 

The hyperuniformity concept can be defined by either the local number variance $\sigma^2(R)$ (i.e., the variance in the number of points within a randomly-thrown spherical window of radius $R$) of the point pattern in the real space or the structure factor $S(\vec{q})$ in the reciprocal or Fourier space \cite{Froufe16}. Here, we use the structure factor $S(\vec{q})$. Hyperuniform point patterns are characterized by a structure factor that vanishes in the long wavelength limit, i.e., $S(q\rightarrow 0)=0$ where $q=|\vec{q}|$, while stealthy hyperuniform point patterns are characterized by a structure factor that vanishes around the origin of wavevectors, $S(q<q_c)=0$ with $\vec{q_c}$ the cut-off reciprocal vector defining the set $\Omega$ in the domain $[0,q_c]$ \cite{Uche04}. 
The distribution of points that meets the conditions on the structure factor can be characterized by the stealthiness $\chi$. The stealthiness is the ratio of the number $M(\Omega)$ of constrained vectors in $\Omega$ to the number of degrees of freedom in the real space $d(N-1)$ in $d$-dimensions (if the system translational degrees of freedom are neglected) \cite{Torquato15}. In this work, we consider a $d=2$-dimensional system and thus $\Omega$ is a circumference of radius $q_c$. Considering the symmetry of the structure factor, $S(\bm{\ks})=S(-\bm{\ks})$ and $M(\Omega)=\frac{1}{2}\pi(q_cL/2\pi)^2$, the stealthiness becomes
\begin{eqnarray}
\chi=\frac{(q_cL)^2}{16\pi (N-1)}.
\label{eq:chi}
\end{eqnarray}
It is worth noting here that this expression of $\chi$ strongly depends on the shape of the domain $\Omega$. Other expressions can be obtained when the $\Omega$ is a square of side $q_c$ as shown in Ref.~\cite{Leseur16}. 

\begin{figure*}
	\centering
\includegraphics[width=16cm]{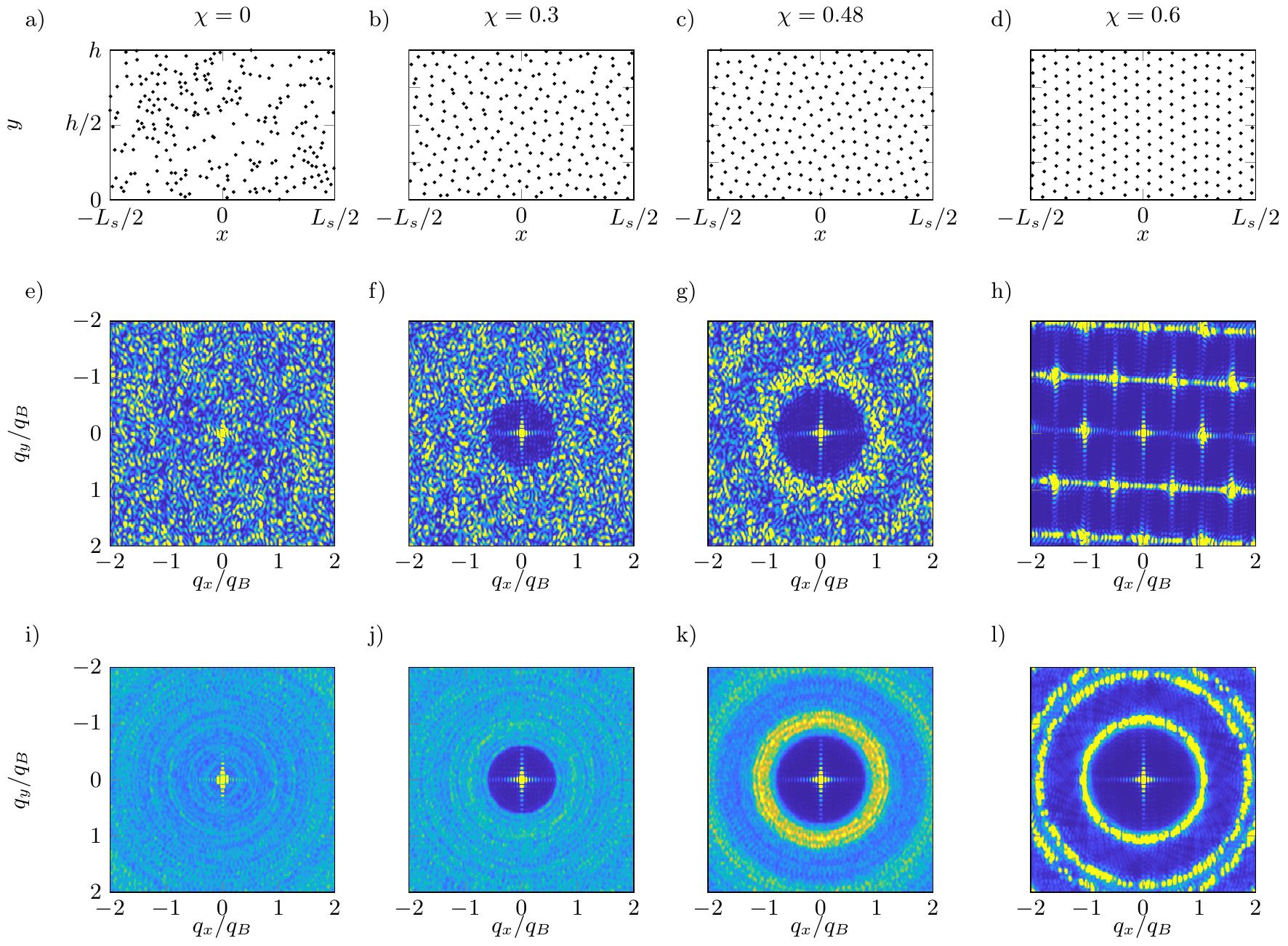}
	\caption{a),b),c),d) Typical configurations of two-dimensional stealthy hyperuniform points pattern scatterers placed in a wave-guide of width $h = 40$ mm and length $\Lslab =1.5h$ for differents values of $\s_factor$. Their corresponding structure factor are presented by the Figs. e),f),i),h) for a single point pattern generation at initial angle $\theta =0$.  i),j),k),l) The corresponding averaged structure factor averaged all over the angle of the media.}
	\label{fig:fig2}
\end{figure*}

A stealthy hyperuniform point pattern characterized by a stealthiness $\chi$ is designed with $N=600$ points embedded in a square area of side $L$. This point pattern is generated using the procedure given by Froufe-Pérez \textit{et al.} \cite{Froufe16}. 
For convenience and because of experimental constrains, we consider a subset of $N_s$ points embedded in a rectangular area of size $L_s\times w$ as shown in Fig. \ref{fig:fig1}(a). Note that other stealthy hyperuniform point patterns with the same $\chi$ can be generated by rotating the initial point pattern by an angle $\theta$ and keeping the points located in the area $L_s\times w$ as shown in Fig. \ref{fig:fig1}(a).

Figure \ref{fig:fig2}(a-d) represents four stealthy hyperuniform point patterns made of $N_s\simeq200$ points, created with the previous procedure, for four different structure factors characterized by four different stealthiness $\chi=[0, 0.3, 0.48, 0.6]$. We have analyzed the evolution of the point patterns by increasing the radius of the constrained area, $q_c$, or equivalently, by increasing the stealthiness $\chi$. The values of the stealthiness are bounded between, $\chi_{min}= 0$ and $\chi_{max}= \pi/4$ (when $q_c=2\pi\sqrt{N}/L$), leading respectively to Poison's distributions and perfect crystal lattices \cite{Torquato15}. The point pattern clearly crystallizes when the stealthiness increases and approaches to 0.5. Figures \ref{fig:fig2}(e-h) show the structure factor of the corresponding point patterns normalized by the amplitude of the Bragg reciprocal vector, $q_B=2\pi\sqrt{N}/L$. The circumference of the constrained area in the reciprocal space of radius $q_c=4\sqrt{\chi N \pi}/L$ is clearly visible. In addition, the structure factor clearly exhibits an extra isotropic region close to $q_c$ for $\chi = 0.48$, where an increase of the structure factor is visible (yellow region). When $\chi\gtrsim 0.5$, the structure factor is anisotropic. The system behaves as a wavy-crystalline system for $\chi=0.6$ as described in Ref.~\cite{Uche04}. Finally, Figs. \ref{fig:fig2}(i-l) show the average structure factor over 60 realizations. The information on the isotropy is lost, but the cutoff wavevector of the hyperuniform materials is clearly visible, where $S(\ks) = 0$ for all $q<q_c$.

\section{Acoustic wave transport in 2D hyperuniform distribution of rigid scatterers}


\subsection{2D Hyperuniform acoustic materials}
\begin{figure}
	\centering
	\includegraphics[width=8cm]{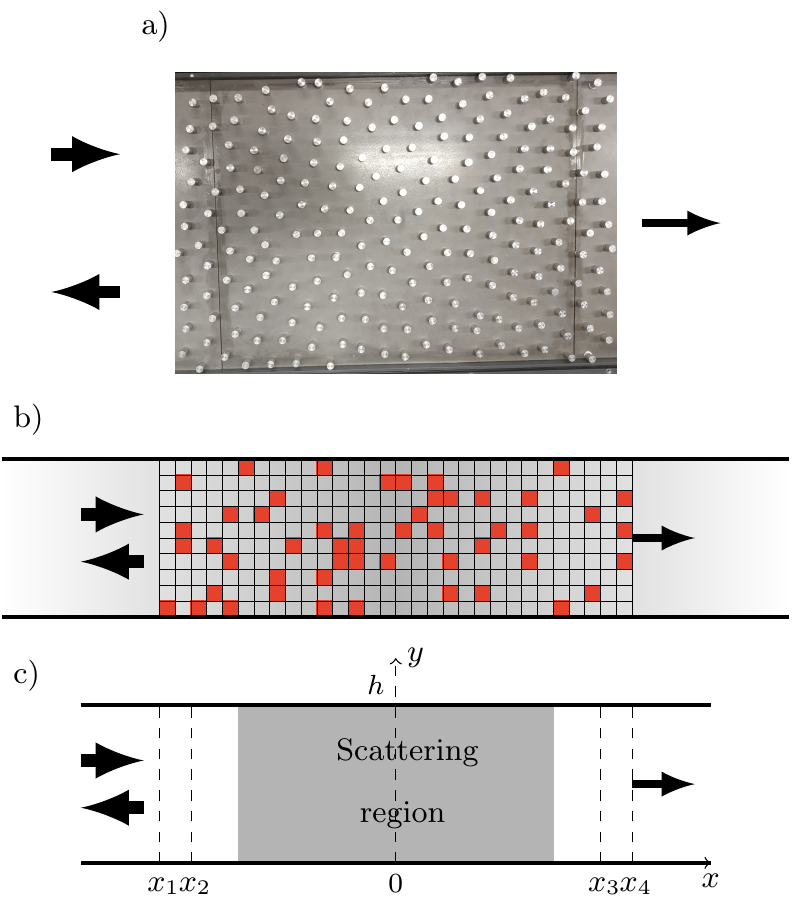}
	\caption{a) Top view of the experimental setup. The scattering region is made of 200 aluminum cylinder of $1$cm diameter. b) Schematics representation of the scattering problem for numerical computation.  c) Schematics representation of the scattering problem with the four lines used to measure pressure field.}
	\label{fig:sketch}
\end{figure}

The hyperuniform material used in this work is made by placing aluminum cylinders of radius ${r_s=0.}5$~cm at the positions of the extracted point pattern in the rectangular area of size ${L_s=0.6}$~m and ${w=0.4}$~m as schematically shown in Fig.~\ref{fig:fig1}(b). Figure~\ref{fig:sketch}(a) shows a picture of one of the 2D hyperuniform acoustic materials experimentally analyzed in this work. 

In order to acoustically characterize the 2D hyperuniform materials, they are embedded in an air filled rectangular waveguide of width $w=0.4$~m and height $h_z=1.5$~cm. The cutoff frequency for the second propagating mode along the waveguide height, i.e., the z-direction, is around 11500 Hz. The studied frequency range is from $400$~Hz to $8000$~Hz, in such a way that the waveguide can be considered as 2D because a single mode can exist along the $z$-direction. The right end of the waveguide is anechoic. The system is excited by a plane wave traveling from the left to the right.

\subsection{Wave transport in complex media embedded in a waveguide}
The wave transport properties of such materials are obtained via their scattering coefficients, i.e., by the reflection and transmission coefficients defined in the next subsection. The wave transport properties of the hyperuniform materials can thus be theoretically, numerically, and experimentally characterized as explained in the following sections. This allows us to analyze the relation between the properties of the point pattern in the reciprocal space with the scattering of the system.

\subsubsection{Multimodal method}

\begin{figure*}
	\centering
	\includegraphics[width=16cm]{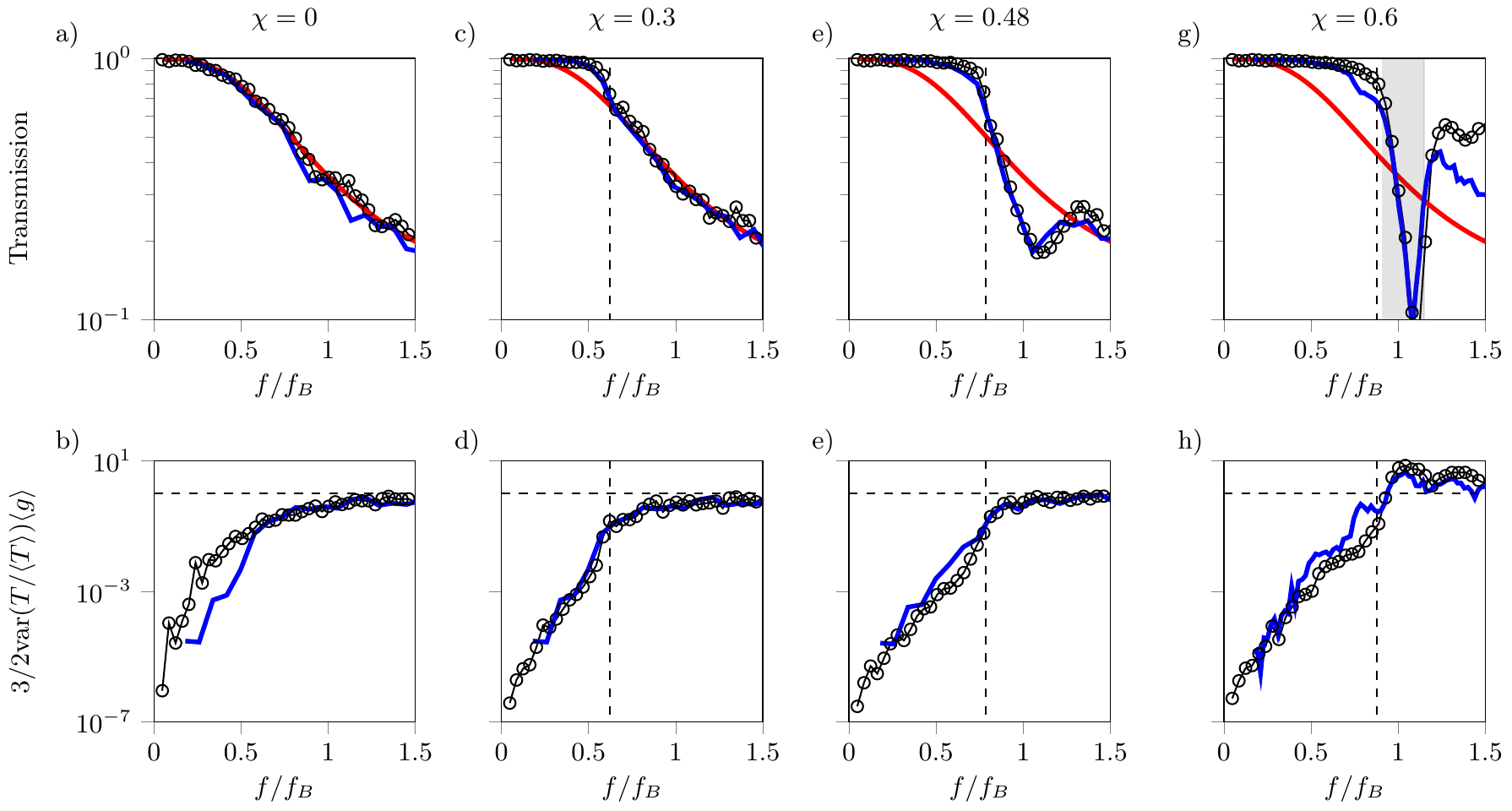}
	\caption{a),c),e),g) Average transmission over 60 iterations through an hypeuniform material as a function of the frequency. The vertical dashed lines represent the cutoff frequency $q_c$ that are determined by Eq.~(\ref{eq:chi}). The red lines are the theoretical transmission deduced from the Ohm's low using the theoretical mean free path of the media (see Supplemental Material). Results using the Multimodal procedure are displayed in blue and results using COMSOL computation are displayed in blue. The band gap of the perfectly crystallized media is highlighted by the grey zone in the $\xi = 0.6$ case (see Supplemental Material).  b),d),e),h) Variance of the transmission as a function of the frequency compared to the theoretical results $3/2\text{var}(\Tg/\avrg{\Tg})\avrg{\g}$ [from Eq.~(\ref{eq:variance})].}
	\label{fig:fig3}
\end{figure*}

The wave transport properties of the stealthy hyperuniform materials are obtained numerically by a multimodal method where the Helmholtz equation ($\nabla^2 p +k^2 p = 0$ with $\partial_n p = 0$ on the rigid boundaries) is projected on the local transverse modes and then solved using an admittance matrix as described in Refs.~\cite{pagneux_multimodal_2010, Maurel2, Maurel1, maurel_modal_2015, felix_sound_2001-1}. This procedure is detailed in the Supplementary material. To solve this problem, we discretize the scattering region on a regular grid of size $\Delta_x = \Delta_y$. The scattering medium of length $L = N_x\Delta_x$ is thus constituted of $N_x$ columns numbered as $i = \lbrace 0,1,...,i,...,N_x\rbrace$ and $N_y$ rows numbered as $j = \lbrace 0,1,...,j,...,N_y\rbrace$. Thus, $w=N_y\Delta_y$ (see Fig.~\ref{fig:sketch}(b)). Here, each column is assumed invariant along the $x$ axis and the associated scattering matrix of the $i$-th column $\bm{S}_i$ is solved using the admittance matrix. The global scattering matrix $\bm{S}$ of the system is calculated by assembling the single scattering matrix of each column characterizing the scattering in the far field region, so accounting for the propagative components. 
The global scattering matrix reads
\begin{eqnarray}
\bm{S} =
\begin{pmatrix}
\bm{R^+}& \bm{T^-}\\ \bm{T^+} & \bm{R^-}
\end{pmatrix},
\end{eqnarray}
where $\bm{R^+}$, $\bm{R^-}$ are the reflection coefficients matrix from each side of the full scattering medium and $\bm{T^+}$ and $\bm{T^-}$ the corresponding transmission matrices. We notice that the system is reciprocal, i.e., $\bm{S}={}^{T}\bm{S}$ ($^T$ meaning transpose) \cite{Pagneux04}.

The general solution for the acoustic pressure, $\psi(x,y)$, can be expressed considering the separation of variables as follows,
\begin{eqnarray}
\psi(x,y)=\sum_m p_m(x)h_m(y),
\end{eqnarray}
where $h_m(y)$, $m \in \mathbb{N}$, is the complete set of orthonormal functions, solutions of the eigenproblem ${h''_m(y)=-k_y^2 h_m(y)}$ (with $h'_m\equiv d h_m/d y$) considering rigid boundary conditions at $y=0$ and $y=h$, i.e., $h_m’(h)=h’_m(0)=0$. The transmission of an incident mode $n$ to a transmitted mode $m$ is given by
\begin{equation}
	p_m^+(L/2) = \sum\limits_nT_{mn} p_n^+(-L/2),
\end{equation}
where $T_{mn}$ are the components of the transmission matrix. The conductance of the system can be calculated directly from these coefficients using the Landauer formula \cite{imry_conductance_1999}
\begin{equation}
	g = \textup{Tr}(\bm{T} \bm{T}^\dagger),
\end{equation}
with $\textup{Tr}()$ the trace and $\bm{T}^\dagger$ the adjoint of $\bm{T}$.
In this work, the average conductance over all angles $\theta \in [0;2\pi]$ is denoted $\avrg{\g}$.
 
We assume that the wave energy is distributed over all modes via multiple scattering process. The sum of the transmission coefficients, corresponding to the transmission of the incident plane wave, is linked to the conductance by
\begin{eqnarray}
\avrg{\Tg} = \avrg{\g}/N_{mod},
\end{eqnarray}
where $N_{mod}$ is the number of propagating modes considered in the solution.

\subsubsection{Experimental and numerical characterization of the scattering properties}

In addition to the multimodal calculations described above, we reconstruct the scattering parameters of the material located in the scattering region as shown in Fig. \ref{fig:sketch}(c). We both experimentally measure and numerically evaluate the pressure field along the $y$ axis of the waveguide at  $41$ equidistant positions. Four lines separated by a distance of $x_2-x_1 = x_4-x_3 = 1.5$~cm upstream ($x_1$, $x_2$) and downstream ($x_3$, $x_4$) are considered to separate both right going and left going waves on both sides of the sample. On the left hand side ($x<-L_s/2$), modes are associated to complex coefficients, $p_n^\pm(x) = c_{n}^\pm e^{\pm ik_nx}$, while on the right hand side ($x>L_s/2$, assuming anechoic termination, i.e., $p_n^-(x) = 0$), modes are only associated to complex coefficients $p_n^+(x)  = d_{n}^+e^{ik_nx}$. We note that in this work the time harmonic convention used is $e^{-i\omega t}$.

The complex coefficients $c^{\pm}_n$ and $d^+_n$ can be obtained via the following approximation of the integral projection on modes $h_n$: 
\begin{equation}
p_n^+(x) + p_n^-(x)\simeq \sum\limits_i \psi(x,y_i)h_n(y_i),
\end{equation} 
where $\psi(x,y_i)$ is the evaluated/measured pressure at the $i$-th position along the $y$ direction of the waveguide at position $x$.
The transmission, through the sample, of an incident plane wave $n=0$ to the $m$-th mode is given by the transmission coefficient
\begin{equation}
	p_m^+(L/2) = T_{m0} p_0^+(-L/2).
\end{equation}

The numerical simulations have been conducted with the acoustic module of COMSOL Multiphysics, considering perfectly matched layers (PML) at the anechoic termination of the waveguide. In the experimental set-up, the anechoic termination is made of a foam block of triangular shape. In the experiments, the plane wave is generated by a set of nine equally-spaced identical high-speakers mounted on the left-end side of the waveguide providing a quasi plane wave excitation along the frequency range of the study.


\subsection{Numerical results in the absence of loss.}

We start the discussion by comparing the transmission coefficients of different stealthy hyperuniform materials with different values of $\chi$, ranging from uncorrelated random to periodic patterns. The transmission coefficients are calculated and compared using  multimodal and finite elements numerical method.
In this section, we do not consider losses, which will be the subject addressed in the next section. 

\begin{figure*}
	\centering
\includegraphics[width=16cm]{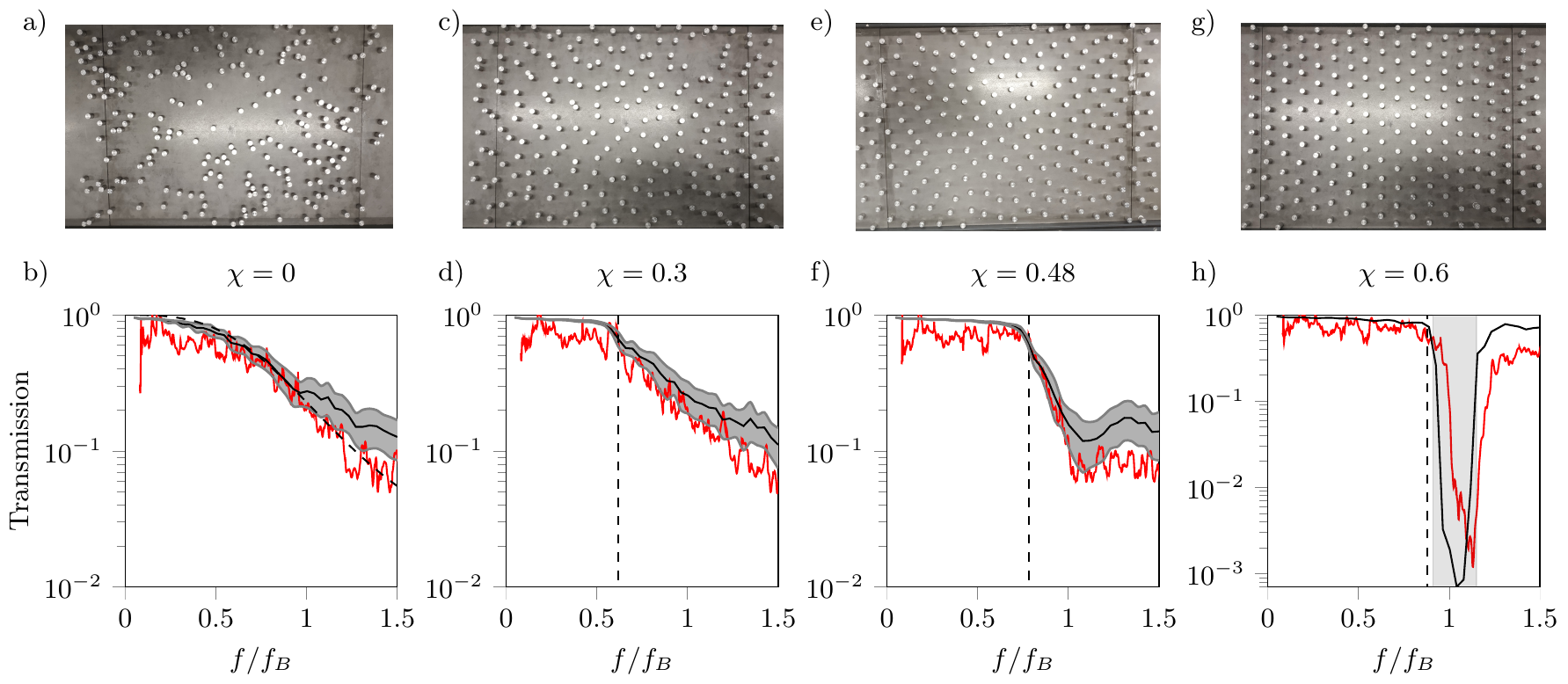}
	\caption{a),c),e),g) Picture of the scatterers distributions used for the experimental study. 
	b),d),f),h) Average transmission through an hyperuniform material as a function of the frequency. The red lines displays the experimental results averaged over 3 configurations. The numerical results averaged over 60 iterations are represented by the black line with a standard deviation displayed by the gray zone. The vertical lines represents the cutoff frequency $\k_c$ determined by Eq. (\ref{eq:chi}). Equation (\ref{eq:brouwer}) for lossy disordered system is represented by the black dotted line for $\chi =0$ using  $\ell_a = 2.5$m as a constant absorption length.  
}
	\label{fig:fig4}
\end{figure*}

The wave transport properties of a disordered material made of a random spatial distribution of scatterers embedded in a waveguide can be described with a single scale parameter $\rL  =  \Lslab/\ell$ \cite{dorokhov_coexistence_1984,mello_macroscopic_1988}. $\ell$ is the transport mean free path that measures the average distance needed for the wave to undergo enough scattering to lose the information of its initial incident direction (see Supplemental Material). We note here that $\ell$ is a frequency dependent parameter. Three main transport regimes are associated with the range of $\rL$ that are accurately described by the Dorokhov-Mello-Pereyra-Kumar (DMPK) equation \cite{dorokhov_coexistence_1984,imry_active_1986,pendry_universality_1992}. When the scattering is weak ($\rL \ll1$), a ballistic transport model can be applied and $\avrg{\g} \approx N_{mod}$. The second regime is the diffusive regime ($1\ll \rL \ll N_{mod}$), where the transverse modes are strongly coupled and the average conductance decreases with the length of the sample according to the Ohm's law 
\begin{eqnarray}
\avrg{\g} = N_{mod}/(1+\rL). 
\end{eqnarray}
The DMPK equation also provides a property related to the variance of the transmission coefficient. In the diffusive regime, the variation is small and does not depend on the scale parameters of the medium. It reads as follows
\begin{equation}
\text{var}(\Tg/\avrg{\Tg})= \dfrac{2}{3}\avrg{\g}.
\label{eq:variance}
\end{equation}
As the diffusion regime is by definition isotropic and produces small variations in the transmission coefficient from a disordered realization to another, we consider this quantity as a threshold to measure the isotropy of the structure. Finally, when $\rL \gg N_{mod}$, the probability of a wave to return to the same coherent volume is not negligible. The interference generated in this volume traps the waves to a finite region of space. The conductance of the medium drops drastically and a transition accours from the diffusive to the localized regimes.

Figure \ref{fig:fig3}(a) shows the average transmission over 60 realizations with $\chi=0$ calculated by both the multimodal method (blue continuous line) and the full wave numerical simulations (black line with open black circles). The predictions by the Ohm's law is shown by the red continuous line. As expected, the three results show the same tendency in agreement with the transport properties of disordered media. In addition to these results, Fig.~\ref{fig:fig3}(b) shows $\dfrac{3}{2}\text{var}(\Tg/\avrg{\Tg})\avrg{\g}$ in terms of frequency.
 As expected, this quantity tends to 1 when the frequency increases. Figures \ref{fig:fig3}(c) and (d) show the corresponding averaged results for 60 realizations with $\chi=0.3$. Vertical line shows the limit $2\sqrt{\chi/N}$ imposed by $\chi$. As previously discussed, the point pattern presents a zero structure factor for $q_x/q_B$ smaller than this value. As shown in Fig.~\ref{fig:fig3}(c), the average transmission is close to 1 in this range of frequencies, showing the characteristic transparency region of stealthy hyperuniform materials. For frequencies higher than this limit, the transmission decreases with the frequency. The average transmission follows the Ohm's law for frequencies higher than the cutoff frequency imposed by $\chi$, i.e. the system behaves as a disordered media in the diffusive regime. Figure \ref{fig:fig3}(d) shows the different behaviors between the transparent and the diffusive regimes which are in agreement with the predictions of the DMPK model, Eq.~(\ref{eq:variance}), in the diffusive regime.

Figure \ref{fig:fig3}(e) and (f) show the average results over 60 realizations with $\chi=0.48$. For this value, the point pattern presents a structure factor with three characteristic regions: the zero region for $q_x<q_c$, an isotropic region with increased values of structure factor [see Fig.~\ref{fig:fig2}(g) and \ref{fig:fig2}(k)], and a region of isotropic random scattering. These three behaviors can be identified for the wave transport properties shown in Fig.~\ref{fig:fig3}(e). For $q_x<q_c$ the transparent region of the stealthy hyperuniform material is shown. Just after the the limit $q_x=q_c$, the transmission presents a dip due to the isotropic region with increased structure factor. In fact, the value of the transmission is smaller than that predicted by the Ohm's law. This is the typical behavior of an isotropic band gap although the point distribution is not periodic. For higher frequencies, the transmission coefficients follows again the Ohm's law, which means that the system falls back into the diffusive behavior. Figure~\ref{fig:fig3}(f) shows the variance, the behavior of which is in accordance with the previous discussion.

Finally, Figs. \ref{fig:fig3}(g) and (h) show the averaged results over 60 realizations with $\chi=0.6$. The point distribution is closer to a periodic pattern. In this case, we can clearly see the transparency region at low frequencies and the presence of the band gap due to the periodicity. The behavior of the system does not follow the Ohm's law meaning that the scattering is anisotropic. The grey area represents the band gap of a triangular lattice as calculated in the Supplementary material. A transmission dip appears in the frequency range of the band gap of the regular lattice showing the hints of periodicity for the structures with $\chi\gtrsim0.5$.

\subsection{Experimental results. Lossy Hyperuniform media}

In this section, we experimentally characterize the wave transport properties of the stealthy hyperuniform materials for airborne sound. For these waves, the losses are unavoidable. They arise from different dissipation mechanisms in complex quasi-1D waveguides that occurs at the scatterer boundaries, in the background medium and at the waveguide walls. Their effects on the wave transport properties of the hyperuniform materials are analyzed. Here, we consider the same configurations as described above accounting for the thermal and viscous losses in the propagation of acoustic waves by simply adding an imaginary part to the wavenumber in the wave equation for the sake of simplicity.

When losses are accounted for in disordered material, the absorption length $\bal$ must be introduced. Two regimes are distinguished depending on the value of this absorption length. The first regime appears when the losses are strong, i.e., when the absorption length is much smaller than the mean free path, i.e., $\bal \ll \tmfp$. In this case, the wave is exponentially damped, $T = \exp{(-L_s/\bal)}$, before the scattering effects appear. The scattering of the wave is negligible and the effect of hyperuniformity and other phenomena arising from scattering are not observed. The second regime appears when the losses are weak, i.e., when the absorption length is much larger than the mean free path, i.e., $\bal \gg \tmfp$. The wave scattering coexists with the losses and an absorbing diffusive transport of the wave takes place.
 
The generalized DMPK equation for disordered systems has been widely studied by Brouwer  \cite{brouwer_transmission_1998} in the absorbing diffusive regime and the transmission can be derived as 
\begin{equation}
	\gbrouwer(\rL,\rdal) = \dfrac{N_{mod}}{\rdal\sinh\left(\dfrac{\rL}{\rdal}\right)+1},
	\label{eq:brouwer}
\end{equation}
where $\rdal \equiv  \xi_a/\tmfp$, with $\xi_a = \sqrt{\bal/2}$ the diffusive absorption length, in mean free path units.

Figures \ref{fig:fig4}(a) and (b) show the image of a stealthy hyperuniform material and its transmission properties respectively for the case $\chi=0$, i.e., for a random distribution of scatterers embedded in the rectangular waveguide. Continuous red line in Fig.~\ref{fig:fig4}(b) shows the experimental transmission averaged over 3 configurations. These results have been used to evaluate the absorption length of the system, which is $\bal=2.5$ m. Black dashed line represents the transmission calculated by the Ohm's law using the conductance considering losses, Eq.~(\ref{eq:brouwer}). Black continuous line shows the numerical results, obtained from the multimodal method, averaged over 60 iterations. The grey area shows the standard deviation. We can observe that the medium presents the behavior predicted by the Ohm's law as for diffusive transport in disordered media.

Figures \ref{fig:fig4}(c) and (d) show the image of a stealthy hyperuniform material and the corresponding average results with $\chi=0.3$. As previously, vertical line shows the limit imposed by $\chi$, i.e., $q_c$. The average transmission represented in Fig.~\ref{fig:fig3}(c) shows the characteristic transparency region of stealthy hyperuniform materials but with an amplitude smaller than the one calculated in the absence of loss, see Fig. \ref{fig:fig3}. This represents a quasi-transparent region. For frequencies higher than the limit $q_c$, we see that the transmission decreases with the frequency. Interestingly, while the average transmission does not follow the Ohm's law in the transparent region, the averaged transmission follows the Ohm's law for higher frequencies than $q_c$, i.e. the system behaves as a disordered media in the absorbing diffusive regime.

Figure \ref{fig:fig4}(e) and (f) show the image of a stealthy hyperuniform material and the corresponding results for with $\chi=0.48$. For this case, the three characteristic regions can be identified even in the presence of losses in Fig.~\ref{fig:fig3}(e). For $q_x<q_c$ the characteristic quasi-transparent region is shown. Just after the limit $q_x=q_c$, a dip of transmission is shown due to the isotropic scattering with increased structure factor. For higher frequencies, the transmission follows the Ohm's law meaning that the system falls back to the diffusive behavior.

Finally, Fig. \ref{fig:fig4}(g) and (h) show the image of a stealthy hyperuniform material and the corresponding results with $\chi=0.6$. In this case, we can clearly see the quasi-transparency region at low frequencies and the presence of the band gap due to the hints of periodicity of the material. We can see that the transmission dip appears in this frequency range coming also from the hints of periodicity for the structures with $\chi\gtrsim0.5$.

\section{Conclusion}
In this work, we have experimentally and numerically analyzed the transport properties of 2D stealthy hyperuniform materials made of rigid scatterer distributions embedded in a waveguide for acoustic waves in the audible regime. The non resonant character of the scatterers allows linking the properties of the structure factor in the reciprocal space with the scattering properties. This shows the presence of tips of transmission by avoiding the need of local resonances. 
The stealthiness $\chi$ imposes a cutoff frequency up to which the structure factor is zero, implying that materials made of rigid scatterer hyperuniform distributions are transparent to waves with frequencies lower than this cut-off frequency. 
These configurations have been also experimentally and numerically analyzed in order to see the feasibility of the structures for the acoustic characterization and for the analysis of the effect of the losses in the wave transport properties. The losses have been accounted for the system via the absorption length that has been phenomenologically recovered from the experiments and used in the theoretical predictions through the generalized DMPK equation. The different regimes discussed previously have been observed even in the presence of the losses. These results open new venues to the control of acoustic waves with disordered materials with target scattering properties.

\begin{acknowledgments}	
This work has been funded by the project HYPERMETA funded under the program \'Etoiles Montantes of the R\'egion Pays de la Loire, by the ANR-RGC METARoom (ANR-18-CE08-0021) project and by the project PID2020-112759GB-I00 of the Ministerio de Ciencia e Innovación.

\end{acknowledgments}


\begin{thebibliography}{46}%
\makeatletter
\providecommand \@ifxundefined [1]{%
 \@ifx{#1\undefined}
}%
\providecommand \@ifnum [1]{%
 \ifnum #1\expandafter \@firstoftwo
 \else \expandafter \@secondoftwo
 \fi
}%
\providecommand \@ifx [1]{%
 \ifx #1\expandafter \@firstoftwo
 \else \expandafter \@secondoftwo
 \fi
}%
\providecommand \natexlab [1]{#1}%
\providecommand \enquote  [1]{``#1''}%
\providecommand \bibnamefont  [1]{#1}%
\providecommand \bibfnamefont [1]{#1}%
\providecommand \citenamefont [1]{#1}%
\providecommand \href@noop [0]{\@secondoftwo}%
\providecommand \href [0]{\begingroup \@sanitize@url \@href}%
\providecommand \@href[1]{\@@startlink{#1}\@@href}%
\providecommand \@@href[1]{\endgroup#1\@@endlink}%
\providecommand \@sanitize@url [0]{\catcode `\\12\catcode `\$12\catcode
  `\&12\catcode `\#12\catcode `\^12\catcode `\_12\catcode `\%12\relax}%
\providecommand \@@startlink[1]{}%
\providecommand \@@endlink[0]{}%
\providecommand \url  [0]{\begingroup\@sanitize@url \@url }%
\providecommand \@url [1]{\endgroup\@href {#1}{\urlprefix }}%
\providecommand \urlprefix  [0]{URL }%
\providecommand \Eprint [0]{\href }%
\providecommand \doibase [0]{http://dx.doi.org/}%
\providecommand \selectlanguage [0]{\@gobble}%
\providecommand \bibinfo  [0]{\@secondoftwo}%
\providecommand \bibfield  [0]{\@secondoftwo}%
\providecommand \translation [1]{[#1]}%
\providecommand \BibitemOpen [0]{}%
\providecommand \bibitemStop [0]{}%
\providecommand \bibitemNoStop [0]{.\EOS\space}%
\providecommand \EOS [0]{\spacefactor3000\relax}%
\providecommand \BibitemShut  [1]{\csname bibitem#1\endcsname}%
\let\auto@bib@innerbib\@empty
\bibitem [{\citenamefont {Schwartz}\ \emph {et~al.}(2007)\citenamefont
  {Schwartz}, \citenamefont {Bartal}, \citenamefont {Fishman},\ and\
  \citenamefont {Segev}}]{Schwartz07}%
  \BibitemOpen
  \bibfield  {author} {\bibinfo {author} {\bibfnamefont {T.}~\bibnamefont
  {Schwartz}}, \bibinfo {author} {\bibfnamefont {G.}~\bibnamefont {Bartal}},
  \bibinfo {author} {\bibfnamefont {S.}~\bibnamefont {Fishman}}, \ and\
  \bibinfo {author} {\bibfnamefont {M.}~\bibnamefont {Segev}},\ }\href
  {\doibase 10.1038/nature05623} {\bibfield  {journal} {\bibinfo  {journal}
  {Nature}\ }\textbf {\bibinfo {volume} {446}},\ \bibinfo {pages} {52}
  (\bibinfo {year} {2007})}\BibitemShut {NoStop}%
\bibitem [{\citenamefont {Wiersma}(2013)}]{Wiersma}%
  \BibitemOpen
  \bibfield  {author} {\bibinfo {author} {\bibfnamefont {D.~S.}\ \bibnamefont
  {Wiersma}},\ }\href@noop {} {\bibfield  {journal} {\bibinfo  {journal} {Nat.
  Photonics}\ }\textbf {\bibinfo {volume} {7}} (\bibinfo {year}
  {2013})}\BibitemShut {NoStop}%
\bibitem [{\citenamefont {Lu}\ \emph {et~al.}(2006)\citenamefont {Lu},
  \citenamefont {Shi}, \citenamefont {Murakowski}, \citenamefont {Schneider},
  \citenamefont {Schuetz},\ and\ \citenamefont {Prather}}]{Lu06}%
  \BibitemOpen
  \bibfield  {author} {\bibinfo {author} {\bibfnamefont {Z.}~\bibnamefont
  {Lu}}, \bibinfo {author} {\bibfnamefont {S.}~\bibnamefont {Shi}}, \bibinfo
  {author} {\bibfnamefont {J.~A.}\ \bibnamefont {Murakowski}}, \bibinfo
  {author} {\bibfnamefont {G.~J.}\ \bibnamefont {Schneider}}, \bibinfo {author}
  {\bibfnamefont {C.~A.}\ \bibnamefont {Schuetz}}, \ and\ \bibinfo {author}
  {\bibfnamefont {D.~W.}\ \bibnamefont {Prather}},\ }\href {\doibase
  10.1103/PhysRevLett.96.173902} {\bibfield  {journal} {\bibinfo  {journal}
  {Phys. Rev. Lett.}\ }\textbf {\bibinfo {volume} {96}},\ \bibinfo {pages}
  {173902} (\bibinfo {year} {2006})}\BibitemShut {NoStop}%
\bibitem [{\citenamefont {Luo}\ \emph {et~al.}(2002)\citenamefont {Luo},
  \citenamefont {Johnson}, \citenamefont {Joannopoulos},\ and\ \citenamefont
  {Pendry}}]{Luo02}%
  \BibitemOpen
  \bibfield  {author} {\bibinfo {author} {\bibfnamefont {C.}~\bibnamefont
  {Luo}}, \bibinfo {author} {\bibfnamefont {S.~G.}\ \bibnamefont {Johnson}},
  \bibinfo {author} {\bibfnamefont {J.~D.}\ \bibnamefont {Joannopoulos}}, \
  and\ \bibinfo {author} {\bibfnamefont {J.~B.}\ \bibnamefont {Pendry}},\
  }\href {\doibase 10.1103/PhysRevB.65.201104} {\bibfield  {journal} {\bibinfo
  {journal} {Phys. Rev. B}\ }\textbf {\bibinfo {volume} {65}},\ \bibinfo
  {pages} {201104} (\bibinfo {year} {2002})}\BibitemShut {NoStop}%
\bibitem [{\citenamefont {Joannopoulos}\ \emph {et~al.}(2008)\citenamefont
  {Joannopoulos}, \citenamefont {Johnson}, \citenamefont {Winn},\ and\
  \citenamefont {Meade}}]{joannopoulos08}%
  \BibitemOpen
  \bibfield  {author} {\bibinfo {author} {\bibfnamefont {J.~D.}\ \bibnamefont
  {Joannopoulos}}, \bibinfo {author} {\bibfnamefont {S.~G.}\ \bibnamefont
  {Johnson}}, \bibinfo {author} {\bibfnamefont {J.~N.}\ \bibnamefont {Winn}}, \
  and\ \bibinfo {author} {\bibfnamefont {R.~D.}\ \bibnamefont {Meade}},\
  }\href@noop {} {\emph {\bibinfo {title} {Photonic Crystals. Molding the Flow
  of Light}}}\ (\bibinfo  {publisher} {Princeton University press},\ \bibinfo
  {year} {2008})\BibitemShut {NoStop}%
\bibitem [{\citenamefont {Deymier}(2013)}]{Deymier13}%
  \BibitemOpen
  \bibinfo {editor} {\bibfnamefont {P.}~\bibnamefont {Deymier}},\ ed.,\
  \href@noop {} {\emph {\bibinfo {title} {Acoustic Metamaterials and Phononic
  Crystals}}}\ (\bibinfo  {publisher} {Springer},\ \bibinfo {year}
  {2013})\BibitemShut {NoStop}%
\bibitem [{\citenamefont {Yablonovitch}(1987)}]{Yablonovitch87}%
  \BibitemOpen
  \bibfield  {author} {\bibinfo {author} {\bibfnamefont {E.}~\bibnamefont
  {Yablonovitch}},\ }\href@noop {} {\bibfield  {journal} {\bibinfo  {journal}
  {Phys. Rev. Lett.}\ }\textbf {\bibinfo {volume} {58}},\ \bibinfo {pages}
  {2059} (\bibinfo {year} {1987})}\BibitemShut {NoStop}%
\bibitem [{\citenamefont {John}(1987)}]{John87}%
  \BibitemOpen
  \bibfield  {author} {\bibinfo {author} {\bibfnamefont {S.}~\bibnamefont
  {John}},\ }\href@noop {} {\bibfield  {journal} {\bibinfo  {journal} {Phys.
  Rev. Lett.}\ }\textbf {\bibinfo {volume} {58 (23)}},\ \bibinfo {pages} {2486}
  (\bibinfo {year} {1987})}\BibitemShut {NoStop}%
\bibitem [{\citenamefont {Sigalas}\ and\ \citenamefont
  {Economou}(1992)}]{Sigalas92}%
  \BibitemOpen
  \bibfield  {author} {\bibinfo {author} {\bibfnamefont {M.}~\bibnamefont
  {Sigalas}}\ and\ \bibinfo {author} {\bibfnamefont {E.}~\bibnamefont
  {Economou}},\ }\href {\doibase https://doi.org/10.1016/0022-460X(92)90059-7}
  {\bibfield  {journal} {\bibinfo  {journal} {Journal of Sound and Vibration}\
  }\textbf {\bibinfo {volume} {158}},\ \bibinfo {pages} {377} (\bibinfo {year}
  {1992})}\BibitemShut {NoStop}%
\bibitem [{\citenamefont {Kushwaha}\ \emph {et~al.}(1993)\citenamefont
  {Kushwaha}, \citenamefont {Halevi}, \citenamefont {Dobrzynski},\ and\
  \citenamefont {Djafari-Rouhani}}]{Kushwaha93}%
  \BibitemOpen
  \bibfield  {author} {\bibinfo {author} {\bibfnamefont {M.~S.}\ \bibnamefont
  {Kushwaha}}, \bibinfo {author} {\bibfnamefont {P.}~\bibnamefont {Halevi}},
  \bibinfo {author} {\bibfnamefont {L.}~\bibnamefont {Dobrzynski}}, \ and\
  \bibinfo {author} {\bibfnamefont {B.}~\bibnamefont {Djafari-Rouhani}},\
  }\href {\doibase 10.1103/PhysRevLett.71.2022} {\bibfield  {journal} {\bibinfo
   {journal} {Phys. Rev. Lett.}\ }\textbf {\bibinfo {volume} {71}},\ \bibinfo
  {pages} {2022} (\bibinfo {year} {1993})}\BibitemShut {NoStop}%
\bibitem [{\citenamefont {Lidorikis}\ \emph {et~al.}(1998)\citenamefont
  {Lidorikis}, \citenamefont {Sigalas}, \citenamefont {Economou},\ and\
  \citenamefont {Soukoulis}}]{Lidorikis98}%
  \BibitemOpen
  \bibfield  {author} {\bibinfo {author} {\bibfnamefont {E.}~\bibnamefont
  {Lidorikis}}, \bibinfo {author} {\bibfnamefont {M.~M.}\ \bibnamefont
  {Sigalas}}, \bibinfo {author} {\bibfnamefont {E.~N.}\ \bibnamefont
  {Economou}}, \ and\ \bibinfo {author} {\bibfnamefont {C.~M.}\ \bibnamefont
  {Soukoulis}},\ }\href {\doibase 10.1103/PhysRevLett.81.1405} {\bibfield
  {journal} {\bibinfo  {journal} {Phys. Rev. Lett.}\ }\textbf {\bibinfo
  {volume} {81}},\ \bibinfo {pages} {1405} (\bibinfo {year}
  {1998})}\BibitemShut {NoStop}%
\bibitem [{\citenamefont {Lidorikis}\ \emph {et~al.}(2000)\citenamefont
  {Lidorikis}, \citenamefont {Sigalas}, \citenamefont {Economou},\ and\
  \citenamefont {Soukoulis}}]{Lidorikis00}%
  \BibitemOpen
  \bibfield  {author} {\bibinfo {author} {\bibfnamefont {E.}~\bibnamefont
  {Lidorikis}}, \bibinfo {author} {\bibfnamefont {M.~M.}\ \bibnamefont
  {Sigalas}}, \bibinfo {author} {\bibfnamefont {E.~N.}\ \bibnamefont
  {Economou}}, \ and\ \bibinfo {author} {\bibfnamefont {C.~M.}\ \bibnamefont
  {Soukoulis}},\ }\href {\doibase 10.1103/PhysRevB.61.13458} {\bibfield
  {journal} {\bibinfo  {journal} {Phys. Rev. B}\ }\textbf {\bibinfo {volume}
  {61}},\ \bibinfo {pages} {13458} (\bibinfo {year} {2000})}\BibitemShut
  {NoStop}%
\bibitem [{\citenamefont {Mie}(1908)}]{Mie08}%
  \BibitemOpen
  \bibfield  {author} {\bibinfo {author} {\bibfnamefont {G.}~\bibnamefont
  {Mie}},\ }\href@noop {} {\bibfield  {journal} {\bibinfo  {journal} {Ann.
  Phys.}\ }\textbf {\bibinfo {volume} {25}} (\bibinfo {year}
  {1908})}\BibitemShut {NoStop}%
\bibitem [{\citenamefont {Bohren}\ and\ \citenamefont
  {Huffman}(1983)}]{Bohren83}%
  \BibitemOpen
  \bibfield  {author} {\bibinfo {author} {\bibfnamefont {C.~F.}\ \bibnamefont
  {Bohren}}\ and\ \bibinfo {author} {\bibfnamefont {D.~R.}\ \bibnamefont
  {Huffman}},\ }\href@noop {} {\emph {\bibinfo {title} {Absorption and
  Scattering of Light by Small Particles}}}\ (\bibinfo  {publisher} {Wiley, New
  York},\ \bibinfo {year} {1983})\BibitemShut {NoStop}%
\bibitem [{\citenamefont {Rockstuhl}\ and\ \citenamefont
  {Lederer}(2006)}]{Rockstuhl06}%
  \BibitemOpen
  \bibfield  {author} {\bibinfo {author} {\bibfnamefont {C.}~\bibnamefont
  {Rockstuhl}}\ and\ \bibinfo {author} {\bibfnamefont {F.}~\bibnamefont
  {Lederer}},\ }\href {\doibase 10.1088/1367-2630/8/9/206} {\bibfield
  {journal} {\bibinfo  {journal} {New Journal of Physics}\ }\textbf {\bibinfo
  {volume} {8}},\ \bibinfo {pages} {206} (\bibinfo {year} {2006})}\BibitemShut
  {NoStop}%
\bibitem [{\citenamefont {Amoah}(2016)}]{Amoah16}%
  \BibitemOpen
  \bibfield  {author} {\bibinfo {author} {\bibfnamefont {T.}~\bibnamefont
  {Amoah}},\ }\emph {\bibinfo {title} {Designer Disordered Complex Media:
  Hyperuniform Photonic and Phononic Band Gap Materials}},\ \href@noop {}
  {Ph.D. thesis},\ \bibinfo  {school} {Advanced Technology Institute and
  Departement of Physics. Faculty of Engineering and Physical Science}
  (\bibinfo {year} {2016})\BibitemShut {NoStop}%
\bibitem [{\citenamefont {Man}\ \emph {et~al.}(2013{\natexlab{a}})\citenamefont
  {Man}, \citenamefont {Florescu}, \citenamefont {Williamson}, \citenamefont
  {He}, \citenamefont {Rez}, \citenamefont {Hashemizad}, \citenamefont {Leung},
  \citenamefont {Liner}, \citenamefont {Torquato}, \citenamefont {Chaikin},\
  and\ \citenamefont {Steinhardt}}]{Man13a}%
  \BibitemOpen
  \bibfield  {author} {\bibinfo {author} {\bibfnamefont {W.}~\bibnamefont
  {Man}}, \bibinfo {author} {\bibfnamefont {M.}~\bibnamefont {Florescu}},
  \bibinfo {author} {\bibfnamefont {E.~P.}\ \bibnamefont {Williamson}},
  \bibinfo {author} {\bibfnamefont {Y.}~\bibnamefont {He}}, \bibinfo {author}
  {\bibfnamefont {S.}~\bibnamefont {Rez}}, \bibinfo {author} {\bibnamefont
  {Hashemizad}}, \bibinfo {author} {\bibfnamefont {B.~Y.~C.}\ \bibnamefont
  {Leung}}, \bibinfo {author} {\bibfnamefont {D.~R.}\ \bibnamefont {Liner}},
  \bibinfo {author} {\bibfnamefont {S.}~\bibnamefont {Torquato}}, \bibinfo
  {author} {\bibfnamefont {P.~M.}\ \bibnamefont {Chaikin}}, \ and\ \bibinfo
  {author} {\bibfnamefont {P.~J.}\ \bibnamefont {Steinhardt}},\ }\href@noop {}
  {\bibfield  {journal} {\bibinfo  {journal} {PNAS}\ }\textbf {\bibinfo
  {volume} {110}},\ \bibinfo {pages} {15886} (\bibinfo {year}
  {2013}{\natexlab{a}})}\BibitemShut {NoStop}%
\bibitem [{\citenamefont {Man}\ \emph {et~al.}(2013{\natexlab{b}})\citenamefont
  {Man}, \citenamefont {Florescu}, \citenamefont {Matsuyama}, \citenamefont
  {Yadak}, \citenamefont {Nahal}, \citenamefont {Hashemizad}, \citenamefont
  {Williamson}, \citenamefont {Steinhardt}, \citenamefont {Torquato},\ and\
  \citenamefont {Chaikin}}]{Man13b}%
  \BibitemOpen
  \bibfield  {author} {\bibinfo {author} {\bibfnamefont {W.}~\bibnamefont
  {Man}}, \bibinfo {author} {\bibfnamefont {M.}~\bibnamefont {Florescu}},
  \bibinfo {author} {\bibfnamefont {K.}~\bibnamefont {Matsuyama}}, \bibinfo
  {author} {\bibfnamefont {P.}~\bibnamefont {Yadak}}, \bibinfo {author}
  {\bibfnamefont {G.}~\bibnamefont {Nahal}}, \bibinfo {author} {\bibfnamefont
  {S.}~\bibnamefont {Hashemizad}}, \bibinfo {author} {\bibfnamefont
  {E.}~\bibnamefont {Williamson}}, \bibinfo {author} {\bibfnamefont
  {P.}~\bibnamefont {Steinhardt}}, \bibinfo {author} {\bibfnamefont
  {S.}~\bibnamefont {Torquato}}, \ and\ \bibinfo {author} {\bibfnamefont
  {P.}~\bibnamefont {Chaikin}},\ }\href@noop {} {\bibfield  {journal} {\bibinfo
   {journal} {Opt. Exp.}\ }\textbf {\bibinfo {volume} {21}},\ \bibinfo {pages}
  {19972} (\bibinfo {year} {2013}{\natexlab{b}})}\BibitemShut {NoStop}%
\bibitem [{\citenamefont {Leseur}\ \emph {et~al.}(2016)\citenamefont {Leseur},
  \citenamefont {Pierrat},\ and\ \citenamefont {Carminati}}]{Leseur16}%
  \BibitemOpen
  \bibfield  {author} {\bibinfo {author} {\bibfnamefont {O.}~\bibnamefont
  {Leseur}}, \bibinfo {author} {\bibfnamefont {R.}~\bibnamefont {Pierrat}}, \
  and\ \bibinfo {author} {\bibfnamefont {R.}~\bibnamefont {Carminati}},\
  }\href@noop {} {\bibfield  {journal} {\bibinfo  {journal} {Optica}\ }\textbf
  {\bibinfo {volume} {3}},\ \bibinfo {pages} {763} (\bibinfo {year}
  {2016})}\BibitemShut {NoStop}%
\bibitem [{\citenamefont {Gkantzounis}\ \emph {et~al.}(2017)\citenamefont
  {Gkantzounis}, \citenamefont {Amoah},\ and\ \citenamefont
  {Florescu}}]{Gkantzounis17}%
  \BibitemOpen
  \bibfield  {author} {\bibinfo {author} {\bibfnamefont {G.}~\bibnamefont
  {Gkantzounis}}, \bibinfo {author} {\bibfnamefont {T.}~\bibnamefont {Amoah}},
  \ and\ \bibinfo {author} {\bibfnamefont {M.}~\bibnamefont {Florescu}},\
  }\href@noop {} {\bibfield  {journal} {\bibinfo  {journal} {Phys. Rev. B}\
  }\textbf {\bibinfo {volume} {95}},\ \bibinfo {pages} {094120} (\bibinfo
  {year} {2017})}\BibitemShut {NoStop}%
\bibitem [{\citenamefont {Aubry}\ \emph {et~al.}(2020)\citenamefont {Aubry},
  \citenamefont {Froufe-P\'erez}, \citenamefont {Kuhl}, \citenamefont
  {Legrand}, \citenamefont {Scheffold},\ and\ \citenamefont
  {Mortessagne}}]{Aubry20}%
  \BibitemOpen
  \bibfield  {author} {\bibinfo {author} {\bibfnamefont {G.~J.}\ \bibnamefont
  {Aubry}}, \bibinfo {author} {\bibfnamefont {L.~S.}\ \bibnamefont
  {Froufe-P\'erez}}, \bibinfo {author} {\bibfnamefont {U.}~\bibnamefont
  {Kuhl}}, \bibinfo {author} {\bibfnamefont {O.}~\bibnamefont {Legrand}},
  \bibinfo {author} {\bibfnamefont {F.}~\bibnamefont {Scheffold}}, \ and\
  \bibinfo {author} {\bibfnamefont {F.}~\bibnamefont {Mortessagne}},\ }\href
  {\doibase 10.1103/PhysRevLett.125.127402} {\bibfield  {journal} {\bibinfo
  {journal} {Phys. Rev. Lett.}\ }\textbf {\bibinfo {volume} {125}},\ \bibinfo
  {pages} {127402} (\bibinfo {year} {2020})}\BibitemShut {NoStop}%
\bibitem [{\citenamefont {Rohfritsch}\ \emph {et~al.}(2020)\citenamefont
  {Rohfritsch}, \citenamefont {Conoir}, \citenamefont {Valier-Brasier},\ and\
  \citenamefont {Marchiano}}]{Rohfritsch20}%
  \BibitemOpen
  \bibfield  {author} {\bibinfo {author} {\bibfnamefont {A.}~\bibnamefont
  {Rohfritsch}}, \bibinfo {author} {\bibfnamefont {J.-M.}\ \bibnamefont
  {Conoir}}, \bibinfo {author} {\bibfnamefont {T.}~\bibnamefont
  {Valier-Brasier}}, \ and\ \bibinfo {author} {\bibfnamefont {R.}~\bibnamefont
  {Marchiano}},\ }\href {\doibase 10.1103/PhysRevE.102.053001} {\bibfield
  {journal} {\bibinfo  {journal} {Phys. Rev. E}\ }\textbf {\bibinfo {volume}
  {102}},\ \bibinfo {pages} {053001} (\bibinfo {year} {2020})}\BibitemShut
  {NoStop}%
\bibitem [{\citenamefont {Romero-Garc{\'\i}a}\ \emph
  {et~al.}(2021)\citenamefont {Romero-Garc{\'\i}a}, \citenamefont {Ch{\'e}ron},
  \citenamefont {Kuznetsova}, \citenamefont {Groby}, \citenamefont {F{\'e}lix},
  \citenamefont {Pagneux},\ and\ \citenamefont {Garcia-Raffi}}]{Romero21}%
  \BibitemOpen
  \bibfield  {author} {\bibinfo {author} {\bibfnamefont {V.}~\bibnamefont
  {Romero-Garc{\'\i}a}}, \bibinfo {author} {\bibfnamefont {{\'E}.}~\bibnamefont
  {Ch{\'e}ron}}, \bibinfo {author} {\bibfnamefont {S.}~\bibnamefont
  {Kuznetsova}}, \bibinfo {author} {\bibfnamefont {J.-P.}\ \bibnamefont
  {Groby}}, \bibinfo {author} {\bibfnamefont {S.}~\bibnamefont {F{\'e}lix}},
  \bibinfo {author} {\bibfnamefont {V.}~\bibnamefont {Pagneux}}, \ and\
  \bibinfo {author} {\bibfnamefont {L.~M.}\ \bibnamefont {Garcia-Raffi}},\
  }\href {\doibase 10.1063/5.0059928} {\bibfield  {journal} {\bibinfo
  {journal} {APL Materials}\ }\textbf {\bibinfo {volume} {9}},\ \bibinfo
  {pages} {101101} (\bibinfo {year} {2021})},\ \Eprint
  {http://arxiv.org/abs/https://doi.org/10.1063/5.0059928}
  {https://doi.org/10.1063/5.0059928} \BibitemShut {NoStop}%
\bibitem [{\citenamefont {Torquato}\ and\ \citenamefont
  {Stillinger}(2003)}]{Torquato03}%
  \BibitemOpen
  \bibfield  {author} {\bibinfo {author} {\bibfnamefont {S.}~\bibnamefont
  {Torquato}}\ and\ \bibinfo {author} {\bibfnamefont {F.~H.}\ \bibnamefont
  {Stillinger}},\ }\href@noop {} {\bibfield  {journal} {\bibinfo  {journal}
  {Phys. Rev. E}\ }\textbf {\bibinfo {volume} {68}} (\bibinfo {year}
  {2003})}\BibitemShut {NoStop}%
\bibitem [{\citenamefont {Uche}\ \emph {et~al.}(2004)\citenamefont {Uche},
  \citenamefont {Stillinger},\ and\ \citenamefont {Torquato}}]{Uche04}%
  \BibitemOpen
  \bibfield  {author} {\bibinfo {author} {\bibfnamefont {O.~U.}\ \bibnamefont
  {Uche}}, \bibinfo {author} {\bibfnamefont {F.~H.}\ \bibnamefont
  {Stillinger}}, \ and\ \bibinfo {author} {\bibfnamefont {S.}~\bibnamefont
  {Torquato}},\ }\href@noop {} {\bibfield  {journal} {\bibinfo  {journal}
  {Phys. Rev. E}\ }\textbf {\bibinfo {volume} {70}},\ \bibinfo {pages} {046122}
  (\bibinfo {year} {2004})}\BibitemShut {NoStop}%
\bibitem [{\citenamefont {Batten}\ \emph {et~al.}(2008)\citenamefont {Batten},
  \citenamefont {Stillinger},\ and\ \citenamefont {Torquato}}]{Batten08}%
  \BibitemOpen
  \bibfield  {author} {\bibinfo {author} {\bibfnamefont {R.~D.}\ \bibnamefont
  {Batten}}, \bibinfo {author} {\bibfnamefont {F.~H.}\ \bibnamefont
  {Stillinger}}, \ and\ \bibinfo {author} {\bibfnamefont {S.}~\bibnamefont
  {Torquato}},\ }\href@noop {} {\bibfield  {journal} {\bibinfo  {journal} {J.
  Appl. Phys}\ }\textbf {\bibinfo {volume} {104}},\ \bibinfo {pages} {033504}
  (\bibinfo {year} {2008})}\BibitemShut {NoStop}%
\bibitem [{\citenamefont {Torquato}\ \emph {et~al.}(2015)\citenamefont
  {Torquato}, \citenamefont {Zhang},\ and\ \citenamefont
  {Stillinger}}]{Torquato15}%
  \BibitemOpen
  \bibfield  {author} {\bibinfo {author} {\bibfnamefont {S.}~\bibnamefont
  {Torquato}}, \bibinfo {author} {\bibfnamefont {G.}~\bibnamefont {Zhang}}, \
  and\ \bibinfo {author} {\bibfnamefont {F.~H.}\ \bibnamefont {Stillinger}},\
  }\href@noop {} {\bibfield  {journal} {\bibinfo  {journal} {Phys. Rev. X}\
  }\textbf {\bibinfo {volume} {5}},\ \bibinfo {pages} {021020} (\bibinfo {year}
  {2015})}\BibitemShut {NoStop}%
\bibitem [{\citenamefont {Torquato}(2016)}]{Torquato16}%
  \BibitemOpen
  \bibfield  {author} {\bibinfo {author} {\bibfnamefont {S.}~\bibnamefont
  {Torquato}},\ }\href@noop {} {\bibfield  {journal} {\bibinfo  {journal}
  {Phys. Rev. E}\ }\textbf {\bibinfo {volume} {94}},\ \bibinfo {pages} {022122}
  (\bibinfo {year} {2016})}\BibitemShut {NoStop}%
\bibitem [{\citenamefont {Froufe-P\'erez}\ \emph {et~al.}(2016)\citenamefont
  {Froufe-P\'erez}, \citenamefont {Engel}, \citenamefont {Damasceno},
  \citenamefont {Muller}, \citenamefont {Haberko}, \citenamefont {Glotzer},\
  and\ \citenamefont {Scheffold}}]{Froufe16}%
  \BibitemOpen
  \bibfield  {author} {\bibinfo {author} {\bibfnamefont {L.~S.}\ \bibnamefont
  {Froufe-P\'erez}}, \bibinfo {author} {\bibfnamefont {M.}~\bibnamefont
  {Engel}}, \bibinfo {author} {\bibfnamefont {P.~F.}\ \bibnamefont
  {Damasceno}}, \bibinfo {author} {\bibfnamefont {N.}~\bibnamefont {Muller}},
  \bibinfo {author} {\bibfnamefont {J.}~\bibnamefont {Haberko}}, \bibinfo
  {author} {\bibfnamefont {S.~C.}\ \bibnamefont {Glotzer}}, \ and\ \bibinfo
  {author} {\bibfnamefont {F.}~\bibnamefont {Scheffold}},\ }\href {\doibase
  10.1103/PhysRevLett.117.053902} {\bibfield  {journal} {\bibinfo  {journal}
  {Phys. Rev. Lett.}\ }\textbf {\bibinfo {volume} {117}},\ \bibinfo {pages}
  {053902} (\bibinfo {year} {2016})}\BibitemShut {NoStop}%
\bibitem [{\citenamefont {Fan}\ \emph {et~al.}(1991)\citenamefont {Fan},
  \citenamefont {Percus}, \citenamefont {Stillinger},\ and\ \citenamefont
  {Stillinger}}]{Fan91}%
  \BibitemOpen
  \bibfield  {author} {\bibinfo {author} {\bibfnamefont {Y.}~\bibnamefont
  {Fan}}, \bibinfo {author} {\bibfnamefont {J.~K.}\ \bibnamefont {Percus}},
  \bibinfo {author} {\bibfnamefont {D.~K.}\ \bibnamefont {Stillinger}}, \ and\
  \bibinfo {author} {\bibfnamefont {F.~H.}\ \bibnamefont {Stillinger}},\
  }\href@noop {} {\bibfield  {journal} {\bibinfo  {journal} {Phys. Rev. A}\
  }\textbf {\bibinfo {volume} {44}},\ \bibinfo {pages} {2394} (\bibinfo {year}
  {1991})}\BibitemShut {NoStop}%
\bibitem [{\citenamefont {Froufe-P{\'e}rez}\ \emph {et~al.}(2017)\citenamefont
  {Froufe-P{\'e}rez}, \citenamefont {Engel}, \citenamefont {S{\'a}enz},\ and\
  \citenamefont {Scheffold}}]{Froufe17}%
  \BibitemOpen
  \bibfield  {author} {\bibinfo {author} {\bibfnamefont {L.~S.}\ \bibnamefont
  {Froufe-P{\'e}rez}}, \bibinfo {author} {\bibfnamefont {M.}~\bibnamefont
  {Engel}}, \bibinfo {author} {\bibfnamefont {J.~J.}\ \bibnamefont
  {S{\'a}enz}}, \ and\ \bibinfo {author} {\bibfnamefont {F.}~\bibnamefont
  {Scheffold}},\ }\href {\doibase 10.1073/pnas.1705130114} {\bibfield
  {journal} {\bibinfo  {journal} {Proceedings of the National Academy of
  Sciences}\ }\textbf {\bibinfo {volume} {114}},\ \bibinfo {pages} {9570}
  (\bibinfo {year} {2017})}\BibitemShut {NoStop}%
\bibitem [{\citenamefont {Bruneau}(2006)}]{bruneau_fundamentals_2006}%
  \BibitemOpen
  \bibfield  {author} {\bibinfo {author} {\bibfnamefont {M.}~\bibnamefont
  {Bruneau}},\ }\href@noop {} {\emph {\bibinfo {title} {Fundamentals of
  acoustics}}}\ (\bibinfo  {publisher} {ISTE Ltd, London; Newport Beach, CA},\
  \bibinfo {year} {2006})\BibitemShut {NoStop}%
\bibitem [{\citenamefont {Kittel}(2004)}]{kittel04}%
  \BibitemOpen
  \bibfield  {author} {\bibinfo {author} {\bibfnamefont {C.}~\bibnamefont
  {Kittel}},\ }\href@noop {} {\emph {\bibinfo {title} {Introduction to Solid
  State Physics}}}\ (\bibinfo  {publisher} {Wiley; 8 edition (November 11,
  2004)},\ \bibinfo {year} {2004})\BibitemShut {NoStop}%
\bibitem [{\citenamefont {Neil W.~Ashcroft}(1976)}]{Ashcroft}%
  \BibitemOpen
  \bibfield  {author} {\bibinfo {author} {\bibfnamefont {N.~D.~M.}\
  \bibnamefont {Neil W.~Ashcroft}},\ }\href@noop {} {\emph {\bibinfo {title}
  {Solid State Physics}}}\ (\bibinfo  {publisher} {Holt, Rinehart and
  Winston},\ \bibinfo {year} {1976})\BibitemShut {NoStop}%
\bibitem [{\citenamefont {Pagneux}(2010)}]{pagneux_multimodal_2010}%
  \BibitemOpen
  \bibfield  {author} {\bibinfo {author} {\bibfnamefont {V.}~\bibnamefont
  {Pagneux}},\ }\href {\doibase 10.1016/j.cam.2009.08.034} {\bibfield
  {journal} {\bibinfo  {journal} {Journal of Computational and Applied
  Mathematics}\ }\textbf {\bibinfo {volume} {234}},\ \bibinfo {pages} {1834}
  (\bibinfo {year} {2010})}\BibitemShut {NoStop}%
\bibitem [{\citenamefont {Maurel}\ and\ \citenamefont
  {Mercier}(2012)}]{Maurel2}%
  \BibitemOpen
  \bibfield  {author} {\bibinfo {author} {\bibfnamefont {A.}~\bibnamefont
  {Maurel}}\ and\ \bibinfo {author} {\bibfnamefont {J.-F.}\ \bibnamefont
  {Mercier}},\ }\href {\doibase 10.1121/1.3682037} {\bibfield  {journal}
  {\bibinfo  {journal} {The Journal of the Acoustical Society of America}\
  }\textbf {\bibinfo {volume} {131}},\ \bibinfo {pages} {1874} (\bibinfo {year}
  {2012})},\ \Eprint {http://arxiv.org/abs/https://doi.org/10.1121/1.3682037}
  {https://doi.org/10.1121/1.3682037} \BibitemShut {NoStop}%
\bibitem [{\citenamefont {Maurel}\ \emph {et~al.}(2014)\citenamefont {Maurel},
  \citenamefont {Mercier},\ and\ \citenamefont {F{\'e}lix}}]{Maurel1}%
  \BibitemOpen
  \bibfield  {author} {\bibinfo {author} {\bibfnamefont {A.}~\bibnamefont
  {Maurel}}, \bibinfo {author} {\bibfnamefont {J.-F.}\ \bibnamefont {Mercier}},
  \ and\ \bibinfo {author} {\bibfnamefont {S.}~\bibnamefont {F{\'e}lix}},\
  }\href {\doibase 10.1121/1.4836075} {\bibfield  {journal} {\bibinfo
  {journal} {The Journal of the Acoustical Society of America}\ }\textbf
  {\bibinfo {volume} {135}},\ \bibinfo {pages} {165} (\bibinfo {year}
  {2014})},\ \Eprint {http://arxiv.org/abs/https://doi.org/10.1121/1.4836075}
  {https://doi.org/10.1121/1.4836075} \BibitemShut {NoStop}%
\bibitem [{\citenamefont {Maurel}\ \emph {et~al.}(2015)\citenamefont {Maurel},
  \citenamefont {Mercier},\ and\ \citenamefont
  {F{\'e}lix}}]{maurel_modal_2015}%
  \BibitemOpen
  \bibfield  {author} {\bibinfo {author} {\bibfnamefont {A.}~\bibnamefont
  {Maurel}}, \bibinfo {author} {\bibfnamefont {J.-F.}\ \bibnamefont {Mercier}},
  \ and\ \bibinfo {author} {\bibfnamefont {S.}~\bibnamefont {F{\'e}lix}},\
  }\href {\doibase 10.1364/JOSAA.32.000979} {\bibfield  {journal} {\bibinfo
  {journal} {Journal of the Optical Society of America A}\ }\textbf {\bibinfo
  {volume} {32}},\ \bibinfo {pages} {979} (\bibinfo {year} {2015})}\BibitemShut
  {NoStop}%
\bibitem [{\citenamefont {F{\'e}lix}\ and\ \citenamefont
  {Pagneux}(2001)}]{felix_sound_2001-1}%
  \BibitemOpen
  \bibfield  {author} {\bibinfo {author} {\bibfnamefont {S.}~\bibnamefont
  {F{\'e}lix}}\ and\ \bibinfo {author} {\bibfnamefont {V.}~\bibnamefont
  {Pagneux}},\ }\href {\doibase 10.1121/1.1391249} {\bibfield  {journal}
  {\bibinfo  {journal} {The Journal of the Acoustical Society of America}\
  }\textbf {\bibinfo {volume} {110}},\ \bibinfo {pages} {1329} (\bibinfo {year}
  {2001})}\BibitemShut {NoStop}%
\bibitem [{\citenamefont {Pagneux}\ and\ \citenamefont
  {Maurel}(2004)}]{Pagneux04}%
  \BibitemOpen
  \bibfield  {author} {\bibinfo {author} {\bibfnamefont {V.}~\bibnamefont
  {Pagneux}}\ and\ \bibinfo {author} {\bibfnamefont {A.}~\bibnamefont
  {Maurel}},\ }\href {\doibase 10.1121/1.1786293} {\bibfield  {journal}
  {\bibinfo  {journal} {The Journal of the Acoustical Society of America}\
  }\textbf {\bibinfo {volume} {116}},\ \bibinfo {pages} {1913} (\bibinfo {year}
  {2004})},\ \Eprint {http://arxiv.org/abs/https://doi.org/10.1121/1.1786293}
  {https://doi.org/10.1121/1.1786293} \BibitemShut {NoStop}%
\bibitem [{\citenamefont {Imry}\ and\ \citenamefont
  {Landauer}(1999)}]{imry_conductance_1999}%
  \BibitemOpen
  \bibfield  {author} {\bibinfo {author} {\bibfnamefont {Y.}~\bibnamefont
  {Imry}}\ and\ \bibinfo {author} {\bibfnamefont {R.}~\bibnamefont
  {Landauer}},\ }\href@noop {} {\bibfield  {journal} {\bibinfo  {journal}
  {Reviews of Modern Physics}\ }\textbf {\bibinfo {volume} {71}},\ \bibinfo
  {pages} {7} (\bibinfo {year} {1999})}\BibitemShut {NoStop}%
\bibitem [{\citenamefont {Dorokhov}(1984)}]{dorokhov_coexistence_1984}%
  \BibitemOpen
  \bibfield  {author} {\bibinfo {author} {\bibfnamefont {O.}~\bibnamefont
  {Dorokhov}},\ }\href {\doibase 10.1016/0038-1098(84)90117-0} {\bibfield
  {journal} {\bibinfo  {journal} {Solid State Communications}\ }\textbf
  {\bibinfo {volume} {51}},\ \bibinfo {pages} {381} (\bibinfo {year}
  {1984})}\BibitemShut {NoStop}%
\bibitem [{\citenamefont {Mello}\ \emph {et~al.}(1988)\citenamefont {Mello},
  \citenamefont {Pereyra},\ and\ \citenamefont
  {Kumar}}]{mello_macroscopic_1988}%
  \BibitemOpen
  \bibfield  {author} {\bibinfo {author} {\bibfnamefont {P.}~\bibnamefont
  {Mello}}, \bibinfo {author} {\bibfnamefont {P.}~\bibnamefont {Pereyra}}, \
  and\ \bibinfo {author} {\bibfnamefont {N.}~\bibnamefont {Kumar}},\ }\href
  {\doibase 10.1016/0003-4916(88)90169-8} {\bibfield  {journal} {\bibinfo
  {journal} {Annals of Physics}\ }\textbf {\bibinfo {volume} {181}},\ \bibinfo
  {pages} {290} (\bibinfo {year} {1988})}\BibitemShut {NoStop}%
\bibitem [{\citenamefont {Imry}(1986)}]{imry_active_1986}%
  \BibitemOpen
  \bibfield  {author} {\bibinfo {author} {\bibfnamefont {Y.}~\bibnamefont
  {Imry}},\ }\href {\doibase 10.1209/0295-5075/1/5/008} {\bibfield  {journal}
  {\bibinfo  {journal} {Europhysics Letters (EPL)}\ }\textbf {\bibinfo {volume}
  {1}},\ \bibinfo {pages} {249} (\bibinfo {year} {1986})}\BibitemShut {NoStop}%
\bibitem [{\citenamefont {Pendry}\ \emph {et~al.}(1992)\citenamefont {Pendry},
  \citenamefont {MacKinnon},\ and\ \citenamefont
  {Roberts}}]{pendry_universality_1992}%
  \BibitemOpen
  \bibfield  {author} {\bibinfo {author} {\bibfnamefont {J.~B.}\ \bibnamefont
  {Pendry}}, \bibinfo {author} {\bibfnamefont {A.}~\bibnamefont {MacKinnon}}, \
  and\ \bibinfo {author} {\bibfnamefont {P.~J.}\ \bibnamefont {Roberts}},\
  }\href {\doibase 10.1098/rspa.1992.0047} {\bibfield  {journal} {\bibinfo
  {journal} {Proceedings of the Royal Society A: Mathematical, Physical and
  Engineering Sciences}\ }\textbf {\bibinfo {volume} {437}},\ \bibinfo {pages}
  {67} (\bibinfo {year} {1992})}\BibitemShut {NoStop}%
\bibitem [{\citenamefont {Brouwer}(1998)}]{brouwer_transmission_1998}%
  \BibitemOpen
  \bibfield  {author} {\bibinfo {author} {\bibfnamefont {P.~W.}\ \bibnamefont
  {Brouwer}},\ }\href {\doibase 10.1103/PhysRevB.57.10526} {\bibfield
  {journal} {\bibinfo  {journal} {Physical Review B}\ }\textbf {\bibinfo
  {volume} {57}},\ \bibinfo {pages} {10526} (\bibinfo {year}
  {1998})}\BibitemShut {NoStop}%
\end{thebibliography}
\end{document}